# All-Dry Transfer of Graphene Film by Van der Waals Interactions

*Seong-Jun Yang,[†] Shinyoung Choi,[†] Francis Okello Odongo Ngome,[‡] Ki-Jeong Kim,[§] Si-Young Choi,[‡] and Cheol-Joo Kim[†,*]*

[†]Department of Chemical Engineering, [‡]Department of Materials Science and Engineering, Pohang University of Science and Technology (POSTECH) and [§]Beamline Research Division, Pohang Accelerator Laboratory (PAL), Pohang, Gyeongbuk 37673, South Korea

ABSTRACT

We report a method that uses van der Waals interactions to transfer continuous, high-quality graphene films from Ge(110) to a different substrate held by hexagonal boron nitride carriers in a clean, dry environment. The transferred films are uniform and continuous with low defect density and few charge puddles. The transfer is effective because of the weak interfacial adhesion energy between graphene and Ge. Based on the minimum strain energy required for the isolation of film, the upper limit of the interfacial adhesion energy is estimated to be 23 meV per carbon atom, which makes graphene/Ge(110) the first as-grown graphene film that has an substrate adhesion energy lower than typical van der Waals interactions between layered materials. Our results suggest that graphene on Ge can serve as an ideal material platform to be integrated with other material systems by a clean assembly process.

Two-dimensional (2d) materials with pristine interfaces form the basis of many functional devices, including transistors, tunneling devices and photodiodes.[1–5] One versatile approach to realize such interfaces is a mechanical transfer technique, by which target materials of interest are mechanically exfoliated from growth substrates by van der Waals interactions with carrier films.[6,7] This approach minimizes atomic defects in the transferred film without changing the in-plane covalent bonds. Additionally, it can preserve clean interfaces of the film by preventing exposure to wet chemicals, which leave ionic and metal impurities.[8] Van der Waals interactions assisted transfer was recently demonstrated with wafer-scale monolayer transition metal dichalcogenides films grown on silicon oxides, and has successfully provided large-scale 2d semiconductors with high quality interfaces.[7] However, application of the technique to another important element, graphene is still limited to micrometer-size samples[6] and fabrication of large-scale graphene devices with pristine interfaces remains challenging.

The difficulty in scaling up the mechanical transfer arises from lack of large-scale graphene samples, which are feasible to exfoliate from the substrates. In particular, the interactions between graphene and carrier film need to overcome the adhesive force of graphene to the underlying substrate. Graphene films grown on typical substrates including Cu and SiC do not satisfy the condition.[9,10] The films adhere to the substrates with adhesion energies ($\gamma$) of 60 meV/carbon (C) atom on Cu,[11] and 106 meV/C atom on SiC,[10] which are significantly higher than typical van der Waals interactions, which have $40 \leq \gamma \leq 50$ meV/C atom between graphene and other layered materials.[12,13] As a result, mechanical transfer by van der Waals forces achieves low yield. To overcome this problem, previous research has used carriers that have stronger adhesions to graphene than to the substrate; the carriers have included polymers that have covalent links, and metals that have strong electrostatic interactions. However, these methods require undesirable wet etching of the carriers.[10,14]

In this report, we present an all-dry, mechanical transfer method for large-scale graphene films by non-destructive van der Waals interactions with hexagonal boron nitride (*h*-BN) carrier films. The key aspect of our developments is to provide easily-detachable graphene films. We achieve the goal by growing graphene on Ge(110) substrates.[15,16] Due to a weak adhesion to the substrate, the graphene film can be effectively exfoliated and transferred, while being attached to a *h*-BN superlayer by van der Waals interactions. The transferred films are uniform and continuous with low density of defects and few charge puddles. The results suggest that our method provides a versatile, clean and scalable way to engineer interfaces of graphene.

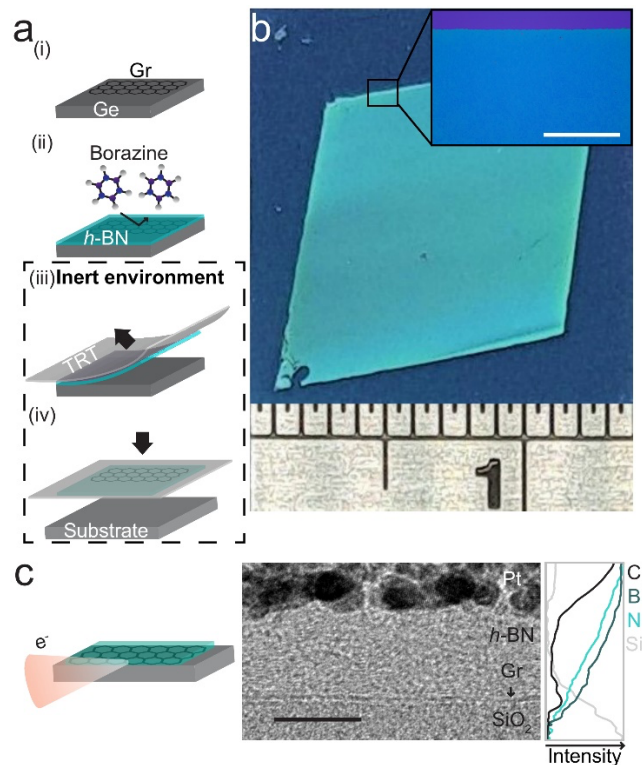

**Figure 1** (a) Schematics of the all-dry transfer process of graphene film by van der Waals Interactions: (i) graphene growth on Ge(110), (ii) *h*-BN growth, (iii) mechanical exfoliation of *h*-BN/graphene hybrid film by TRT and (iv) transfer of the film to arbitrary substrates. (b) Optical image of 50 nm *h*-BN/graphene film transferred onto a SiO$_2$/Si substrate. Scale bar: 100 μm. (c) (left) Schematics for cross-sectional TEM, (middle) TEM image and (right) EDX intensity profile for carbon, boron, nitrogen and silicon elements. Scale bar: 10 nm.

Our transfer method (Figure 1a; Supporting Information) entails four main steps: (i) a high-quality graphene film with uniform thickness is grown on a Ge(110) substrate by chemical vapor deposition (CVD), (ii) a *h*-BN carrier film is subsequently grown on the top surface of graphene, (iii) the whole *h*-BN/graphene composite film is mechanically exfoliated using a thermal-release tape (TRT), and (iv) transferred onto a target substrate in an inert environment.

We chose Ge(110) substrate and *h*-BN carrier materials to ensure high transfer yield and cleanness at the graphene interfaces. Ge(110) is a good template to grow a weakly-bonded graphene film for feasible exfoliations, because it has a dissimilar lattice structure to graphene and low free carrier concentration to suppress electrostatic interactions.[17] Adhesion of graphene is significantly weaker to Ge(110) than to other growth substrates.[16] Here, we test whether the adhesion is even weaker than van der Waals interactions between graphene and other layered materials; if so, the weak adhesion would enable mechanical transfer assisted by van der Waals interaction. For that purpose, we grew *h*-BN films on graphene and used them as carriers. The surface of *h*-BN has no dangling bonds, so it has minimal effect on the electrical properties of graphene. Therefore, we do not need to remove *h*-BN, so interfaces of graphene remain clean.

We first checked the feasibility of the transfer by optically examining the films at macroscopic scales. An optical image (Figure 1b) of a centimeter-sized *h*-BN/graphene film transferred onto a 300-nm silicon dioxide/silicon ($SiO_2$/Si) substrate shows uniform optical contrast over the whole area, compared to the bare $SiO_2$/Si substrate; this uniformity demonstrates transfer of a continuous, homogeneous film without wrinkles or cracks. In order to identify the contents of the transferred film, energy-dispersive X-ray (EDX) elemental map was obtained by transmission electron microscopy (TEM) in a 50 nm thin slide of 10 nm *h*-BN/graphene on a $SiO_2$/Si substrate, that was carved out using a focused ion beam. The EDX intensity profiles (Fig. 1c) for each element including carbon, boron, nitrogen and silicon across the interfaces clearly identify both *h*-BN and graphene layers on the SiO₂ surface. Also, optical absorption spectra were measured with a beam diameter of 0.5 cm in a film transferred onto a fused silica substrate (Supporting Information, Fig. S1) and the presences of both *h*-BN and graphene films were confirmed over the large-area by their optical spectra signatures.

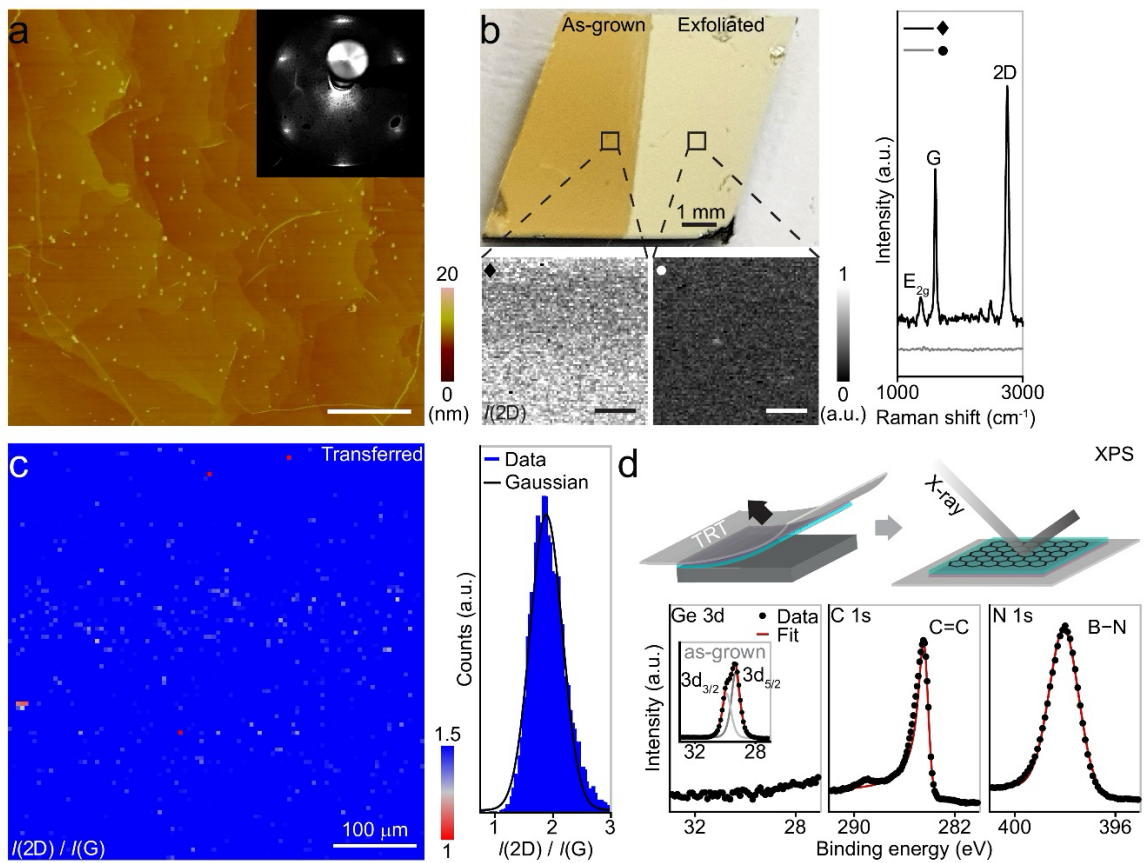

**Figure 2** (a) (main panel) AFM height image and (inset) LEED pattern measured from a graphene film grown on a Ge(110) substrate. Scale bar: 1 μm. (b) Photograph of a 45 nm *h*-BN/graphene hybrid film on a Ge(110) substrate, where only the right side of the film is exfoliated for comparison. Left-bottom images: Raman intensity maps for graphene 2D peak measured from the squared regions of (diamond) the as-grown and (circle) exfoliated parts. Right panel: Each Raman spectra. Scale bars: 10 μm. (c) (left) 2D/G Raman peak intensity ratio map of 20 nm *h*-BN/graphene film transferred onto a SiO₂/Si substrate and (right) distribution of the ratios. (d) (top) Schematics for XPS measurements and (bottom: left to right) XPS spectra for Ge 3d, C 1s and N1s peaks.

The transfer yield was confirmed by using spatially-resolved characterizations after each stage of the processes (Fig. 1a) to observing the structural homogeneity of the films. Our method achieved almost 100 % transfer of as-grown graphene films with ultra-flat surfaces and low density of atomic defects, while preserving excellent structural homogeneity with uniform monolayer thickness and low density of atomic defects throughout the processes. First, atomic force microscopy (AFM) height image of an as-grown graphene sample (Fig. 2a) shows an ultra-flat surface with a few terrace structures of atomic steps. While nanoscale particles are observed in the film, their formations can be suppressed, if the Ge substrate undergoes a pre-cleaning process before the growth of graphene. (Supporting Information, Fig. S2) Low energy electron diffraction (LEED) pattern (Fig. 2a, inset) and Raman spectra image with 488 nm excitation wavelength (Supporting Information, Fig. S3) confirmed the uniform monolayer thickness and the low density of atomic defects. The high-quality samples were provided for subsequent CVD of *h*-BN carrier films. An optical image (Fig. 2b) of a *h*-BN/graphene film on a Ge(110) substrate was obtained after exfoliation of only the right side of the sample (Supporting Information, Fig. S4). Raman spectra of the as-grown region (Fig. 2b, right panel) shows clear G and 2D peaks of graphene and the $E_{2g}$ peak of *h*-BN (1373 cm$^{-1}$), but not in spectra of the exfoliated region. Raman intensity maps of graphene 2D peak (Fig. 2b, bottom panel) further demonstrate effective exfoliation of the film. In the as-grown region, the 2D peak intensity shows a variation of ± 33 % of the average value over the scanned region, possibly due to the inhomogeneous strain on graphene film.[17] Nonetheless, we found that the subsequent *h*-BN growth does not introduce significant graphene Raman D peak associated with atomic defects, if the initial graphene film has a uniformly high Raman 2D intensity. (Supporting Information, Fig. S5 and S6) Finally, we transferred the film onto a SiO$_2$/Si substrate and obtained a 2D/G Raman peak intensity ratio map (Fig. 2c). The 2D/G intensity ratio is spatially uniform; the distribution has a single Gaussian peak with an average value of 1.9 (Fig. 2c, right panel); this result indicates that the graphene was a uniform monolayer (Supporting Information, Fig. S7).[18]

To check whether the transfer process induces contamination to graphene, we examined the bottom surface of exfoliated graphene film by using X-ray photoelectron spectroscopy (XPS) (Fig. 2d, upper schematics). Apparent Ge 3d peaks from the as-grown sample (Fig. 2d, left-most inset) are totally absent from graphene after its isolation from a Ge substrate (Fig. 2d, left-most spectra). Over the full energy range of XPS measurement from 0 to 1300 eV, signals associated with metal elements are absent; this result indicates that the surface is free of metal impurities. (Supporting Information, Fig. S8). In contrast, C 1s peaks are mostly preserved; the main peak at 284.4 eV is assigned to carbons that have a graphitic $sp^2$ bond.[19,20] Furthermore, the cross-sectional TEM image on a transferred film (Fig. 1c) doesn't show contamination bubbles with few tens of nanometers in size at the interfaces, which are evidently observed in abundance in samples that are transferred by a method involving wet processes to etch polymer and metallic supporting films.[8,21] While the possibility of airborne contaminations at an atomic level cannot be completely ruled out in the current stage,[22] the lack of hydrocarbon and metal-related features in TEM and XPS measurements suggests that contaminations are substantially reduced at nanometer scales by all-dry transfer method. Randomly distributed ionic or metal impurities cause inhomogeneous doping effects and result in non-uniform electrical properties of graphene devices. These effects hamper realizations of novel phenomena, and applications of graphene devices.[4,6,23] Our transfer method can eliminate these limitations by providing clean and defect-free graphene channels for electronic devices, as demonstrated below.

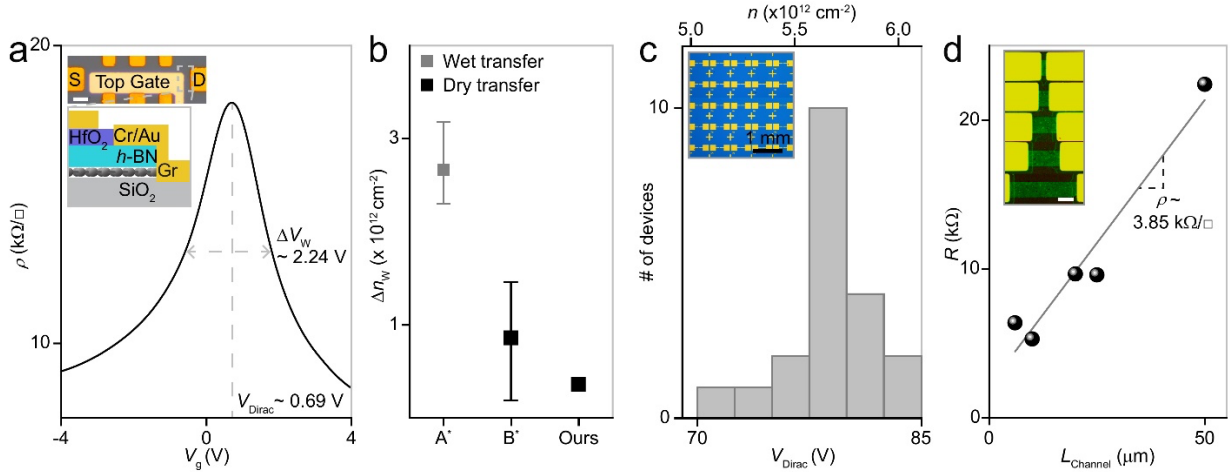

**Figure 3** (a) Top gate voltage-dependent sheet resistivity of a 65 nm $h$-BN/graphene channel on a fused silica substrate. Inset: optical image and a schematic for the device. Scale bar: 10 μm. (b) Inhomogeneous charge variations for different graphene channels, deduced from Dirac peak widths based on a capacitance model. A[*] and B[*] are extracted from Ref. 26 and 25 respectively. (c) Distribution of Dirac voltages and corresponding doping levels measured at 50 nm $h$-BN/graphene channel arrays on a SiO$_2$/Si substrate. (inset) (d) Channel length dependent resistances of 50 nm $h$-BN/graphene channels with different lengths. All the graphene samples are prepared by the all-dry transfer method. Scale bar: 10 μm.

We fabricated (Supporting Information) field-effect transistors (FETs) based on graphene films transferred by the all-dry method, then evaluated their electrical properties. The graphene channels were protected by a $h$-BN superlayer throughout the fabrication processes and only the edges of graphene were exposed to make electrical contacts.[6] The structure and dielectric properties of $h$-BN films, which are used as a protective layer and a gate dielectric layer are characterized by TEM, AFM and capacitance measurements. (Supporting Information) Consequently, $h$-BN films show amorphous like structures with nanometer size domains (Fig. S9) with the dielectric constant of 2 and the breakdown field of 0.5 MV/cm (Fig. S10), similar to the properties of previously reported multilayer films grown by chemical vapor deposition.[24] For a $h$-BN/graphene channel on a fused silica (Fig. 3a, inset), sheet resistivity ($\rho$) at room temperature varied as a function of the top gate voltage ($V_g$). By measuring the full width at half maximum ($\Delta V_w$) of $\rho$ near the charge-neutral point, we estimate the upper limit of inhomogeneous charge puddle ($\Delta n_w$) at the low doping level by using a simple capacitance model $\Delta n_w = \Delta V_w C_g/e$, where $C_g$ is the gate capacitance per unit area and $e$ is the elementary charge. The estimated value is $\Delta n_w = 3.6 \times 10^{11}$ cm$^{-2}$, which is comparable to the smallest values measured by the same method from small flakes on SiO$_2$/Si substrates,[25] and much smaller than the values from wet-transferred films (Fig. 3b).[26]

We also measured transconductance of an array of FETs with back-gated structures on a SiO$_2$/Si substrate (Fig. 3c) by a batch fabrication. On an SiO$_2$/Si substrate, the graphene channels show typical *p*-type doping characteristics with the charge-neutral $V_g$ ($V_{Dirac}$) at high positive voltages, similar to previous reports. Our measured $V_{Dirac}$ had a small variation within $10^{12}$ cm$^{-2}$, which is comparable to the doping inhomogeneity induced by charge puddles in clean graphene films on SiO$_2$/Si substrates.[27] The observation suggests that the doping homogeneity is mainly determined by the surface structure rather than by contamination caused by the transfer process, and that homogeneity in the sample is maintained from micrometer scale to centimeter scale. As a result, graphene channels with different lengths have well-defined resistivity at zero $V_g$ (Fig. 3d). The electrical properties, especially the carrier mobility, can still be improved (Supporting Information, Fig. S11). Nevertheless, our results demonstrate that this all-dry transfer process based on graphene/Ge(110) provides uniform films without additional disorders.[4]

The effective transfer indicates that adhesion energy between graphene and Ge(110) ($\gamma_{Gr-Ge}$) is indeed weaker than the van der Waals interactions between graphene and *h*-BN interfaces. Here, we further measure $\gamma_{Gr-Ge}$ quantitatively to provide a useful guideline to evaluate the feasibility of graphene transfer by other materials beside *h*-BN. To estimate $\gamma_{Gr-Ge}$ we measured the strain energy ($E_\varepsilon$) required to exfoliate graphene films from Ge substrates. When *h*-BN superlayer is deposited on graphene to more than a critical thickness of 50 nm, the *h*-BN/graphene films spontaneously delaminated, with buckling (Supporting Information, Fig. S12). This phenomenon typically

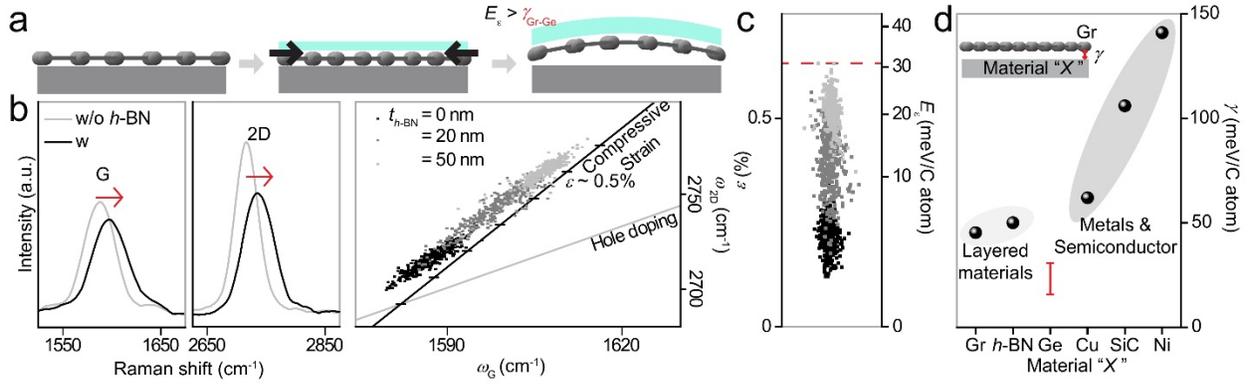

**Figure 4** (a) Schematics of buckling of graphene film induced by compressive strain by superlayer deposition. When residual strain energy ($E_\varepsilon$) exceeds the interfacial adhesion energy ($\gamma_{Gr-Ge}$) between graphene and the substrate, spontaneous buckling can occur. (b) left: Representative Raman spectra for graphene G and 2D peaks before and after h-BN depositions. The excitation wavelength is 488 nm; right: 2D peak positions ($\omega_{2D}$) versus G peak positions ($\omega_G$) measured at multiple spots of samples with different h-BN thicknesses. Overlain lines: compressive strain (black) and hole doping (gray) dependent $\omega_{2D}$ and $\omega_G$ taken from previous studies.[29,30] (c) Compressive strain and corresponding $E_\varepsilon$ of graphene films from Fig. 4b, deduced by $\omega_{2D}$ and $\omega_G$. Red dotted line: maximum value of strain and $E_\varepsilon$; *i.e.*, at critical thickness. (d) Interfacial adhesion energy between graphene and other materials from previous reports (graphene[12], h-BN[13], Cu[11], SiC[10] and Ni[11]). $\gamma_{Gr-Ge}$ is deduced from the maximum $E_\varepsilon$ in Fig. 4c.

happens when compressive residual $E_\varepsilon$ exceeds $\gamma_{Gr-Ge}$ (schematics in Fig. 4a).[28] By using micro Raman spectroscopy, we tried to monitor the accumulative compressive strain in the graphene films as h-BN superlayers thickened, until the strain was released by the spontaneous delamination. Raman spectra (Figure 4b) near G and 2D peaks of graphene measured before and after deposition of h-BN superlayers show blue shifts in both peaks after the deposition; the shifts increased as the thickness of h-BN film increased (Fig. 4b, right panel). We compared our data with previous reports, where the peak shifts are measured as a function of two dominant factors for phonon stiffening of graphene, *i.e.*, compressive strain and hole doping (Fig. 4b, right panel).[29,30] 2D peak position ($\omega_{2D}$) and G peak position ($\omega_G$) changed in response to compressive strain. The compressive strain was deconvoluted using a simple equation based on linear combinations of peak shifts caused by the two dominant factors; the deduced strain shows a maximum value of about 0.67 % in the samples with *h*-BN films above the critical thickness.

Based on the maximum strain values measured from multiple samples, we deduce the maximum $E_\varepsilon$ in *h*-BN/graphene film prior to the delamination. According to fracture mechanics,[28,31] the $E_\varepsilon$ is considered to be equivalent to $\gamma_{Gr-Ge}$, when all $E_\varepsilon$ is applied to overcome $\gamma_{Gr-Ge}$ for the delamination. $E_\varepsilon = \varepsilon^2 E_{Gr} t_{Gr}/(1-\nu_{Gr})$, where $\varepsilon$, $E_{Gr}$, $t_{Gr}$ and $\nu_{Gr}$ are respectively the strain, Young's modulus, thickness, and Poisson ratio of graphene, if most of the residual strain is accumulated at graphene, which is significantly softer than the much thicker *h*-BN superlayer. The calculation yielded $\gamma_{Gr-Ge}$ = 23 ± 7.5 meV/C atom (Fig. 4d). This value is considered as the upper limit of $\gamma_{Gr-Ge}$. In reality, not all of $E_\varepsilon$ is used for the delamination, so a residual stress remains in the final film (Supporting Information, Fig. S12). The calculated $\gamma_{Gr-Ge}$ is the smallest ever reported for interfacial adhesion energies between an as-grown graphene film and a growth substrate.[10,11] The value is even smaller than the $\gamma$ of van der Waals interactions between layered materials; this conclusion is consistent with the high yield transfer of graphene by interaction with *h*-BN superlayer.[13]

The feasibly detachable graphene on Ge(110) further enables a general "*contact and peel*" approach (Fig. 5a) to produce transferable graphene films with engineered top interfaces. In detail, one can stack arbitrary material *X* other than *h*-BN on the top surface of as-grown graphene film, either by deposition or by transfer of *X* to form clean *X*/graphene interfaces. Then the composite films can be peeled and transferred onto arbitrary substrates to exploit the properties of *X*/graphene interfaces. The interfaces are formed in an inert environment, so pristine interfaces should be obtained without oxidation of *X*. Moreover, the whole film can be transferred onto arbitrary substrates, so the properties of *X*/graphene composites can be studied without any effects from the growth substrate. This approach can be applied to fabricate various interfaces at a high production yield as long as $\gamma_{Gr-Ge}$ is weakest among the all interlayer adhesions in the stacked composite films before exfoliation, and facilitates fabrication of devices by integrating other material components.

By applying the "*contact and peel*" approach to as-grown graphene film on Ge(110), we fabricated (Supporting Information) transferable graphene films interfaced with arbitrary material $X$ ($X$ = Cu, Ni, *h*-BN and graphene), then transferred the films onto target substrates (Fig. 5b, inset). Raman spectra of all samples (Fig. 5b) show clear G and 2D peaks of single-layer graphene, confirming successful formation and transfer of $X$/graphene interfaces. In particular, we can fully isolate graphene film on Ge(110) by stacking another graphene film on top, then exfoliating the stacked double-layer graphene to realize layer-by-layer assembly of 2d materials. In this approach, as the individual graphene films have aligned crystalline structure (Fig. 2a, inset), twisted bilayer graphene (tBLG) with a controlled interlayer rotational angle ($\theta_{stack}$) can be fabricated. (Fig. 5c, left panel) As an example, tBLG with $\theta_{stack}$ = 15° is fabricated and the real part of optical conductivity ($\sigma$) is deduced from optical transmission measurements.[32] The measured $\sigma$ from tBLG ($\sigma_{tBLG}$) shows twice the value of $\sigma$ from single-layer graphene ($\sigma_{SLG}$) in most of the spectral range, proving effective formation and transfer of tBLG. When the value of $\sigma_{tBLG}$ spectra (Fig. 5c, black) is subtracted by twice the value of $\sigma_{SLG}$ spectra (Fig. 5c, gray), an extra absorption peak (Fig. 5c, orange) is observed at 2.56 eV. This energy is close to the inter-band transition energy of 2.6 eV between van Hove singularities resulting from the interlayer interactions between the two graphene layers with $\theta$ = 15°. In order to check the interlayer coupling efficiency, we compared the integrated absorption peak intensity with the value measured from as-grown tBLG with clean interfaces.[32] The integrated absorption peak intensity is comparable to the reference value, suggesting a formation of clean interfaces with efficient interlayer interactions. (Supporting Information, Fig. S13)

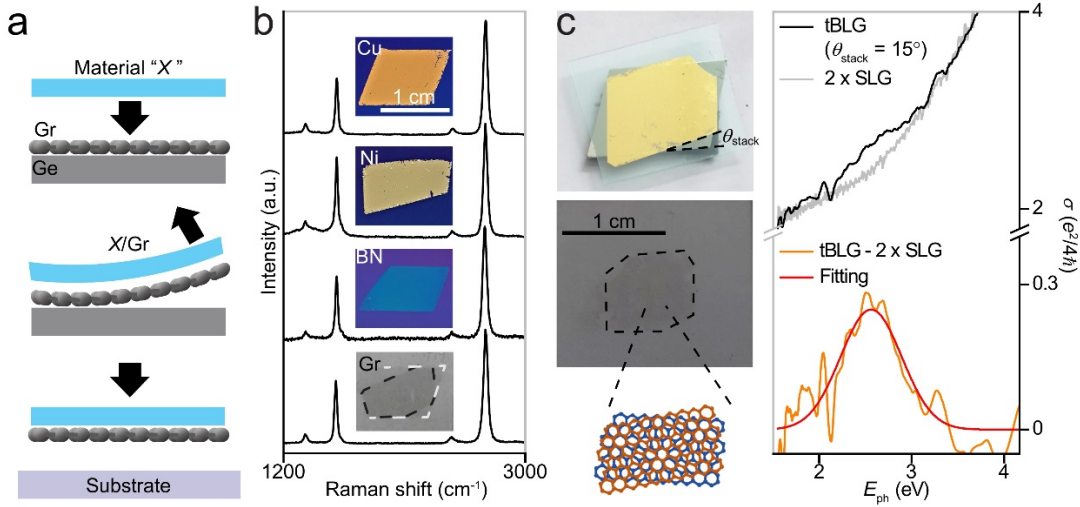

**Figure 5** (a) Schematics for a "*contact and peel*" process to fabricate graphene films interfaced with arbitrary material, *X*. (b) (inset, top to bottom) Photographs of *X*/graphene films (*X* = Cu, Ni, *h*-BN and graphene) transferred onto $SiO_2$/Si or fused silica substrates and (main panel) corresponding Raman spectra. Metal superlayers were chemically removed for the Raman measurements. (c) (left, from top to bottom) Photograph of stacked tBLG with $\theta_{stack}$ = 15° on Ge, transferred tBLG on fused silica and schematics of tBLG. (right) Measured optical conductivity ($\sigma$) spectra of tBLG with $\theta_{stack}$ = 15° ($\sigma_{tBLG}$, black), 2 × $\sigma$ spectra of single-layer graphene (2 × $\sigma_{SLG}$, gray) and $\sigma_{tBLG}$ − 2 × $\sigma_{SLG}$ (orange).

In summary, we report that graphene films grown on Ge(110) are transferable by van der Waals interactions with supporting superlayers in an all-dry environment. Our results demonstrate that graphene films on Ge(110) can serve as an ideal component for a versatile assembly process to integrate graphene with other material systems.[33,34] In particular, recent studies have shown that Ge can also serve as a growth substrate for other layered materials. For example, Ge(100) and Ge(110) can facilitate the growth of semiconducting graphene nanoribbons and insulating *h*-BN monolayer on their surfaces. Apparently, the adhesion energies of the materials to the underlying Ge are weak, at least below 60 meV/C atom.[35,36] The weak adhesion implies that our all-dry transfer method on Ge substrate can be applied to the atomically thin layered materials, enabling the assembly of complex structures with more building blocks. The approach will be useful for realizing large-scale 2d electronics, and for discovering new hybrid materials that have novel electrical and optical properties.[37–40]

ASSOCIATED CONTENT

**Supporting Information**. The Supporting Information includes following contents.

Experimental methods for CVD of graphene on Ge(110) substrate, all-dry transfer process of graphene film, TEM sample fabrication and characterization, dielectric property characterization of *h*-BN films, device fabrication and "*contact and peel*" process. Supporting data for optical absorption spectra of transferred *h*-BN/graphene hybrid film, AFM height images of graphene film on Ge with different chemical treatments, Raman spectra of graphene film on Ge(110), all-dry transfer process of graphene film, Raman spectra change in the defective graphene film after *h*-BN growth, Raman spectra change in the high-quality graphene film after *h*-BN growth, Raman spectra of multi-layer graphene film, X-ray photoelectron spectroscopy on an exfoliated graphene surface, structural characterization of *h*-BN films, dielectric properties of *h*-BN films, field-effect carrier mobility of graphene channels, calculation of interfacial toughness between graphene and Ge(110) based on the geometry of formed buckles and interlayer optical absorption peak area in artificially fabricated tBLG

AUTHOR INFORMATION

**Corresponding Author**

*E-mail: kimcj@postech.ac.kr

**Notes**

The authors declare no competing financial interest.

ACKNOWLEDGMENT

We thank J. Park for experimental help. This research was supported by the International Research & Development Program (NRF-2017K1A3A1A12073407) and the Creative Materials Discovery Program (NRF-2018M3D1A1058793) of the National Research Foundation of Korea (NRF) funded by the Ministry of Science, ICT & Future Planning, as well as the Basic Science Research Program through the NRF funded by the Ministry of Education (NRF-2017R1D1A1B03034896). Experiments at PLS-II were supported in part by MSICT and POSTECH. Additional funding was provided by POSTECH Basic Science Research Institute. SYC is supported by the Global Frontier R&D Program on Center for Hybrid Interface Materials (HIM) funded by the Ministry of Science, ICT & Future Planning Korea (NRF-2013M3A6B1078872).

# Experimental Methods

**Chemical vapor deposition of graphene on Ge(110) substrate**

To grow graphene film, we conduct ambient pressure CVD on Ge(110) substrate by using a recipe modified from methods reported elsewhere.[1,2] First, sonicate undoped Ge(110) substrate (Item #: GEUe100D050C1-US, MTI Corporation) for 5 min in acetone and isopropyl alcohol (IPA) to remove organic residues, then dip into a $H_2O_2 + H_2O + H_3PO_4$ solution (6 : 3 : 1 ratio in volume) under heating at 110 °C for 1 h to etch the top surface of Ge. Sonicate Ge substrate in $H_2O$ for 5 min to remove particles and dip into 20 % aqueous HF solution for 5 min to etch native oxide. Load Ge(110) substrate in a hot-wall, quart tube furnace. Evacuate the furnace to 200 mTorr, then fill it with Ar (200 sccm) and $H_2$ (20 sccm), until it reaches ambient pressure. Ramp up the temperature to 915 °C (ramping rate: 45 °C / min) with a flow of Ar (200 sccm) and $H_2$ (10 sccm) and anneal the substrate for 20 min at 915 °C. Flow 1 % $CH_4$ diluted in $H_2$ (8 sccm) and Ar (200 sccm) for 4.5 h. Reduce the temperature to 600 °C with the same gas flow, then further cool it to room temperature with Ar (200 sccm) and $H_2$ (10 sccm).

For the earlier growths including the one for the sample, which is shown in Fig. 2a of the main manuscript, Ge etching process with a $H_2O_2 + H_2O + H_3PO_4$ solution was not involved. In the case, randomly distributed nanoscale particles were occasionally observed in the film. (Fig. S2a) The particles could be removed by a HF oxide etchant, indicating that they were Ge oxides. Later, their formations were effectively suppressed by treating the Ge substrate with a $H_2O_2 + H_2O + H_3PO_4$ solution before the growth of graphene. (Fig. S2b)

**All-dry transfer process of graphene film**

For the all-dry transfer of graphene, as-grown graphene films are subsequently followed by growth of h-BN carrier films, then exfoliation and transfer. We flow Ar carrier gas through liquid borazine (Product code #: INBO009, Gelest) with high vapor pressure to supply sufficient precursor into the reaction chamber to grow multilayer h-BN films.[3] The temperature of precursor is kept at -20 °C. The experimental details for each process are described below.

*Growth of h-BN carrier film*

Load as-grown graphene on Ge(110) substrate in a hot-wall, quart tube furnace. Evacuate the furnace to 200 mTorr, then fill it with Ar (200 sccm) and $H_2$ (10 sccm) until it reaches ambient pressure. Ramp up the temperature to 720 °C (ramping rate: 45 °C / min) with a flow of Ar (200 sccm) and $H_2$ (10 sccm). Start to flow borazine precursor (5 sccm) in addition to the base gases (Ar (200 sccm) and $H_2$ (10 sccm)) at 350 °C during the ramping up. Grow h-BN film to a desired thickness at 720 °C. The growth rate is ~ 0.4 nm/min. Reduce the temperature to room temperature with Ar (200 sccm) $H_2$ (10 sccm).

In terms of formations of atomic defects, the effect of the subsequent h-BN growth on the quality of the underlying graphene strongly depends on the initial quality of graphene and the growth temperature of h-BN films. In order to check the effect of h-BN growth on graphene quality, we perform Raman spectra mapping on a same region of graphene before and after the process of h-BN deposition. Before the optimization of graphene growth, there are regions, where the 2D peak intensity is significantly lower than the majority of the area, as indicated by a diamond symbol in Fig. S5. For quantitative analysis, the 2D peak intensity ($I$(2D)) is normalized by Raman peak intensity ($I$(Ge)) near 300 cm$^{-1}$ associated with Ge-Ge vibrations. As a result, in the region, where the $I$(2D) / $I$(Ge) is below 0.08, graphene D peaks associated with atomic defects significantly increase after *h*-BN growth, while in the rest of the areas, such increment of D peaks is not observed. We effectively suppress the formation of additional defects in graphene during the *h*-BN growth by an optimized growth of graphene, which eliminates the lower $I$(2D) / $I$(Ge) (below 0.08) regions.

Also, it is important to maintain the growth temperature of *h*-BN film below the damage threshold for graphene. Here, the temperature for *h*-BN growth is maintained below 750 °C, above which even graphene with high $I$(2D) / $I$(Ge) values starts to develop D peaks after *h*-BN CVD. To further confirm the effect of subsequent *h*-BN growth on graphene quality, $I$(2D) / $I$(G) Raman intensity ratio map is compared in a high-quality graphene film before and after *h*-BN growth. (Fig. S6) There is a certain reduction of the $I$(2D) / $I$(G) ratio, but the values are maintained above the value, 1 after *h*-BN growth, indicating that the crystalline structures of graphene film are preserved. We currently attribute the reduction of the ratio to the residual strains on graphene, that is induced by *h*-BN superlayer, but further investigation is required to fully understand the phenomena.

*Exfoliation and transfer of h-BN/graphene film*

Spin coat PMMA (996 K, 8 % in anisole) on *h*-BN/graphene film at 3000 rpm for 50 s and attach a thermal-release tape (TRT) to the sample. Load the whole substrate inside a glove box in Ar environment. Exfoliate the film by pulling out TRT, then attach it to a desired substrate. Release the TRT by heating the substrate at 135 °C on a hot plate. Remove PMMA by annealing at 350 °C for 2 h.

**TEM sample fabrication and characterization**

For cross sectional TEM analysis, we use a standard focused-ion beam lift-out to prepare a cross-section of *h*-BN/graphene on a SiO$_2$/Si substrate. Imaging and energy-dispersive X-ray (EDX) chemical mapping are conducted with a 60 kV operation voltage. After acquiring EDX spectrum images in the *h*-BN/graphene interface area, we extract chemical composition depth profile of boron, nitrogen, carbon and silicon across the interfaces by integrating the intensities over the whole scanned area. (main manuscript, Fig. 1c)

**Dielectric property characterization of *h*-BN films**

For the characterization of dielectric property of *h*-BN films, we fabricate a metallic disk (90 nm Au / 10 nm Cr film with 680 μm$^2$ area) on the top of 60-nm-thick *h*-BN/Gr films grown on Ge, and measure the breakdown field across the dielectric *h*-BN film between the top disk and the bottom graphene. (Fig. S10) Breakdown field of 0.5 MV/cm is defined at a current density of ~10$^{-2}$ A cm$^{-2}$, which is just above the noise floor of the measurement. Small capacitance (*C*) is measured with a 200 mV a.c. input signal of 1 kHz frequency with zero d.c. bias on metal /30 nm *h*-BN/ metal structures, and *h*-BN dielectric constant ($\varepsilon_r$) of 2 is deduced, using $C = \varepsilon_r \varepsilon_0 A / t_{BN}$, where $\varepsilon_0$, $A$ and $t_{BN}$ are the vacuum dielectric permittivity, capacitance area and the thickness of *h*-BN film.

**Device fabrication**

Dry-transferred graphene films are used to fabricate field effect transistors. Photolithography is performed to pattern a protective layer of photoresist on the transferred films. Reactive ion-etching ($CF_4$ 40 sccm / $O_2$ 4 sccm, Power 50 W) is used to etch the unprotected *h*-BN/graphene and define the geometry of graphene channels.[4,5] Then 10 nm Cr / 50 nm Au is thermally evaporated to form electrical contacts. Specifically, the sample is metallized by 40° tilting to expose the edges of graphene channels toward the evaporated sources. To make a metallic contact on the other side edge, the sample is tilted toward a different direction accordingly. For a device with a top gate, we deposit an aluminium layer of 1 nm, and oxide it by annealing at 100 °C on a hot plate to use it as a seed layer to deposit 30 nm of dielectric $HfO_2$ layer by atomic layer deposition. Top gate electrodes with 2 nm Cr / 30 nm Au are formed by a lithography and metallization process.

**"*Contact and peel*" process**

The "*contact and peel*" process provides transferable films with desired interfaces between graphene and arbitrary material *X*. If the material *X* (e.g., Cu, Ni and Au) can be evaporated, we deposit it on top of as-grown graphene/Ge(110) under a high vacuum condition in a thermal evaporator. The deposition rate is usually 0.1 ~ 0.4 Å/s for the first 10 nm, then 1 Å/s for the remaining thickness. If the superlayer induced strain energy is close to the interfacial adhesion energy between graphene and Ge(110), entire films can be easily exfoliated by peeling it off.[6] The critical thickness for the top film to induce a sufficient strain is a few tens of nanometers.

To make graphene/graphene interfaces, we first mechanically exfoliate the 1$^{st}$ graphene film after deposition of carrier film (usually copper) and attachment of TRT. We place this film on top of the 2$^{nd}$ graphene film right after it is unloaded from the growth chamber. After a gentle pressure is applied, the whole film is exfoliated again by TRT and transferred onto a target substrate. The carrier film is etched away, if necessary.

# Supporting Information

**Optical absorption spectra of transferred *h*-BN/graphene hybrid film**

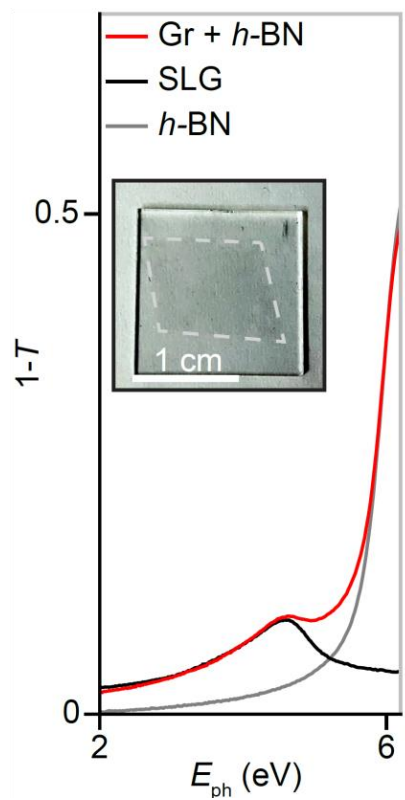

**Figure S1** (a) Optical spectra of 1 – transmission (*T*) equivalent to absorption, measured from *h*-BN/graphene hybrid film (red), single-layer graphene (SLG, black) and 20 nm *h*-BN (gray) on a fused silica. Inset: photograph of *h*-BN/graphene hybrid film on a fused silica substrate. The measured 1 - *T* spectrum closely matches the sum of 1 – *T* spectra of single-layer graphene and 20-nm-thick *h*-BN, which are measured separately from individual samples. This result demonstrates that both films are present.

**AFM height images of graphene film on Ge with different chemical treatments**

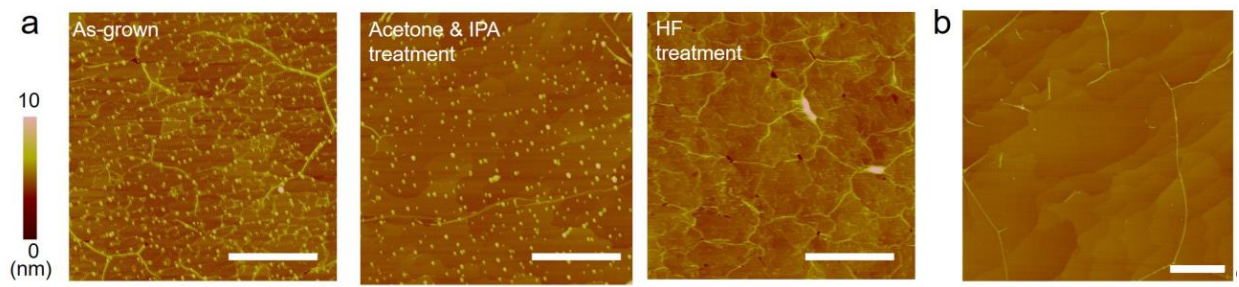

**Figure S2** (a) (from left to right) AFM height images of as-grown graphene film on Ge, acetone + IPA-treated sample and HF-treated sample. Nanoscale particles are removed after HF treatment, implying that the particles are inorganic oxides. (b) AFM height image of graphene film grown on the Ge substrate after a pre-treatment process by a Ge and GeO$_x$ etchant (H$_2$O$_2$ + H$_2$O + H$_3$PO$_4$ solution) for 1 h with heating at 110 °C. The nanoscale particles are absent after graphene growth, if the Ge surface is pre-treated by the etchant before the growth. Scale bars: 1 μm.

**Raman spectra of graphene film on Ge(110)**

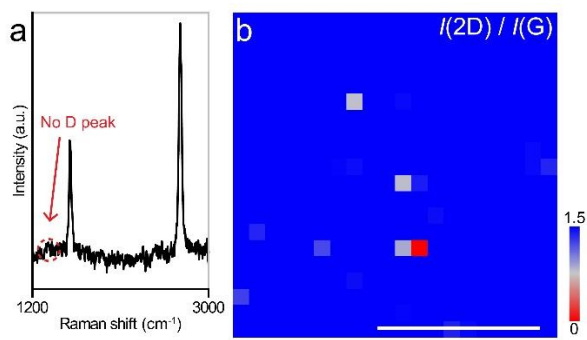

**Figure S3** (a) Representative Raman spectra and (b) $I(2D) / I(G)$ Raman peak ratio map of as-grown graphene film on Ge(110) substrate (excitation wavelength = 488 nm). The defect-related phonon energy of 1350 cm$^{-1}$ is absent; *i.e.*, the film had low defect density. The mapping data show $I(2D) / I(G) > 1.5$ over 98 % of the scanned area; this result confirms uniform monolayer thickness. Scale bar: 10 μm.

**All-dry transfer process of graphene film**

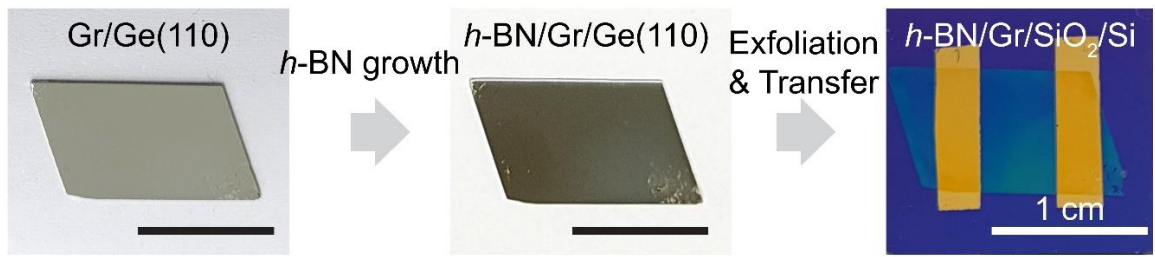

**Figure S4** Photographs of graphene samples after each step of the all-dry transfer process.

**Raman spectra change in the defective graphene film after *h*-BN growth**

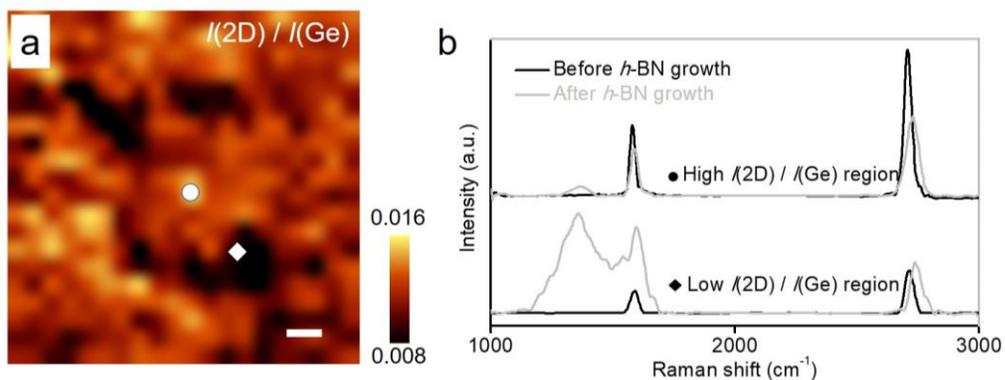

**Figure S5** (a) Raman intensity maps for graphene 2D peak normalized by the Ge peak at 300 cm$^{-1}$ (($I$(2D) / $I$(Ge)) before *h*-BN growth. Scale bar: 1 μm. (b) Raman spectra of graphene film before and after *h*-BN growth in a high $I$(2D) / $I$(Ge) area (top spectra) and a low $I$(2D) / $I$(Ge) area (bottom spectra), which are indicated by circle and diamond symbols in Fig. S5a, respectively. Atomic defect associated D peak significantly increases only in the areas with $I$(2D) / $I$(Ge) < 0.08.

**Raman spectra change in the high-quality graphene film after *h*-BN growth**

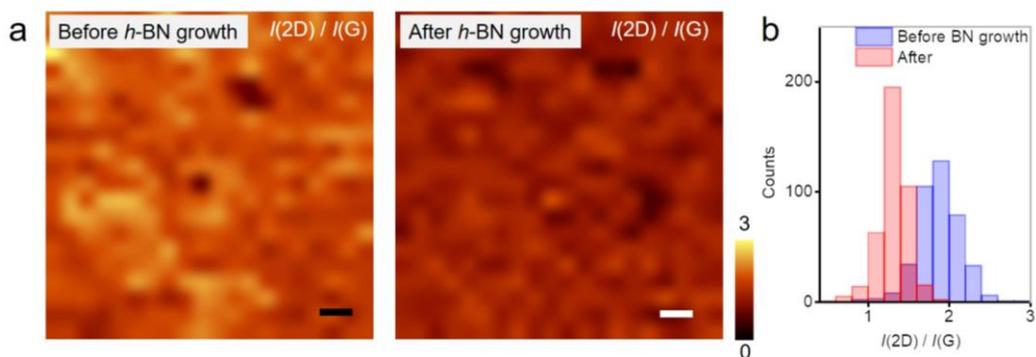

**Figure S6** (a) 2D/G Raman peak intensity ratio maps of graphene film before and after growth of *h*-BN film on top. Scale bars: 1 μm. (b) Histogram of 2D/G Raman peak intensity ratio.

**Raman spectra of multi-layer graphene film**

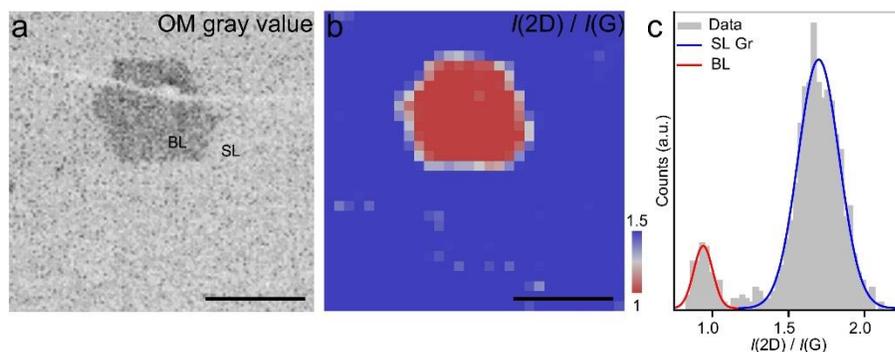

**Figure S7** (a) Optical image and (b) $I$(2D) / $I$(G) Raman peak ratio map of graphene film on SiO$_2$/Si substrate with single-layer (SL) and bi-layer (BL) regions. Scale bars: 10 μm. (c) Distribution of $I$(2D) / $I$(G) from Fig. S7b. The ratio from SL and BL regions of graphene show clearly-distinguished distributions with average values of 1.7 for SL graphene and 0.94 for BL graphene by Gaussian fitting.

**X-ray photoelectron spectroscopy on an exfoliated graphene surface**

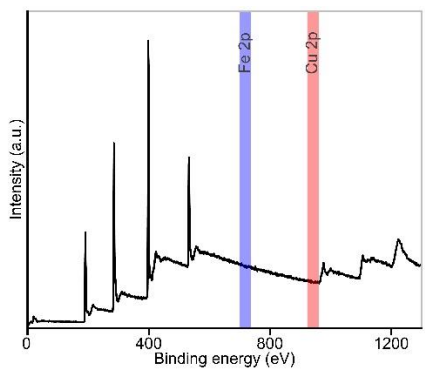

**Figure S8** Full X-ray photoelectron spectroscopy (XPS) data of exfoliated graphene surface from Fig. 2d in the main manuscript. No peak appears in the regions of Fe 2p (blue) or Cu 2p (red) binding energy.

## Structural characterization of *h*-BN Films

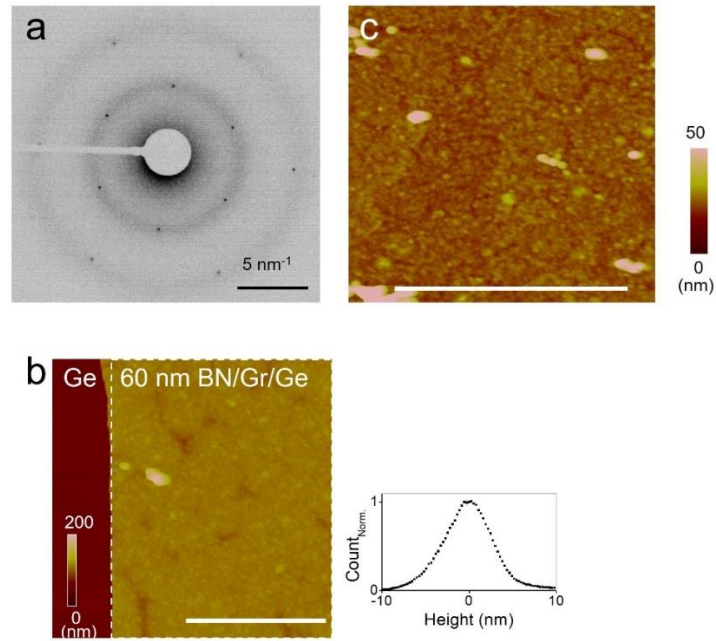

**Figure S9** (a) TEM diffraction pattern from a selected area (140 nm in the radius) of a 3 nm *h*-BN/graphene film. As the underlying graphene has an aligned crystalline structure (main manuscript, Fig. 2a, inset), the hexagonal diffraction spots and the ring pattern are considered to be originated from graphene, and *h*-BN superlayer, respectively, indicating that the *h*-BN film has amorphous like structures. (b) (left) Height image of an as-grown *h*-BN/graphene film of 60 nm thickness (right) Histogram of the height distribution from the dotted square region of 2 × 1.6 μm$^2$ in the left image. The surface roughness is measured as 2.8 nm. (c) AFM height image of a transferred 3 nm *h*-BN/graphene film on a SiO$_2$/Si substrate. The surface roughness is measured as 2.4 nm. Scale bars: 1 μm.

## Dielectric properties of *h*-BN films

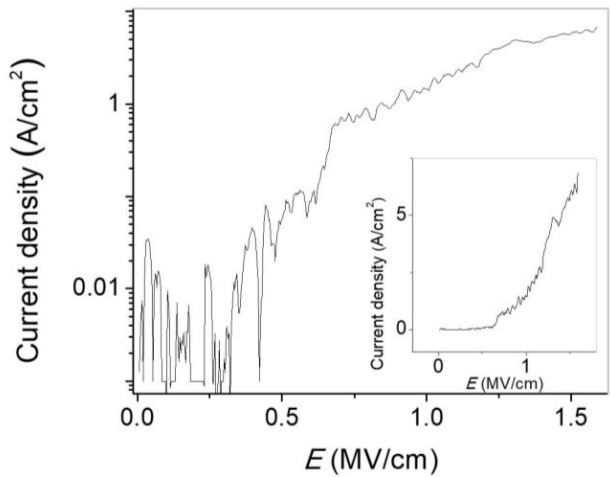

**Figure S10** (main) Current density versus applied electric field for a 60 nm *h*-BN film. (electrode area: 680 μm$^2$) Breakdown field is defined at a current density of ~$10^{-2}$ A cm$^{-2}$, which is just above the noise floor of the measurement.

**Field-effect carrier mobility of graphene channels**

Field-effect carrier mobility is measured in two probe graphene channels on a fused silica substrate transferred by the all-dry method. Conductivity ($\sigma$) changed as a function of top-gate induced carrier concentration ($n$) (Figure S11) which is deduced by a simple capacitance model. Based on a self-consistent Boltzmann model for diffusive transport, we fit the data using $\sigma^{-1} = (ne\mu_c + \sigma_0)^{-1} + \rho_s$, where $\mu_c$, $\sigma_0$ and $\rho_s$ are carrier concentration-independent mobility due to long-range Coulomb scattering, residual conductivity at the charge neutral point and resistivity determined by short-range scattering.[7] As a result, $\mu_c$, $\sigma_0$ and $\rho_s$ are estimated to be 2180 cm$^2$/V·s, 2 $e^2/h$ and 5430 Ω/□ for hole carriers, respectively. The carrier mobility is similar to previously-reported values in graphene grown on Ge(110), but lower than the best values for CVD graphene grown on metallic substrates.[8] The difference is mainly a result of the high $\rho_s$ in our samples, which is caused by atomic defects such as grain boundaries. We expect that $\mu_c$ can be increased by optimizing the growth process.

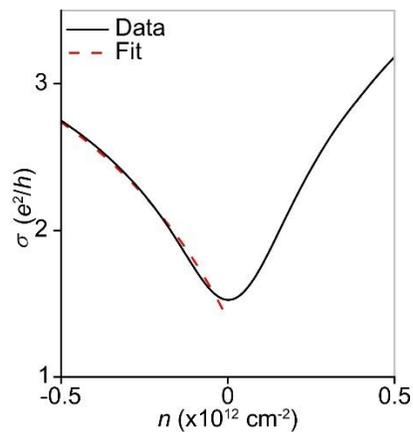

**Figure S11** Transconductance as a function of carrier concentration induced at the top gate. Dotted red line: fitting based on a self-consistent Boltzmann model for diffusive transport.

**Calculation of interfacial toughness between graphene and Ge(110) based on the geometry of formed buckles**

Interfacial fracture energy of a particular interface is commonly measured by observing change of geometry induced by a stressed overlayer. If the induced compressive stress exceeds a critical limit ($\sigma_c$), the film begins to buckle, and driving stress ($\sigma_d$, $\sigma_d > \sigma_c$) induces further propagation of the buckles. $\sigma_c$ and $\sigma_d$ are related to the geometry of the buckles and to the mechanical properties of the film:[9]

$$\sigma_c = \frac{\pi^2 E}{12(1-v^2)}\left(\frac{t}{w/2}\right)^2, \quad \text{(Eq. 1)}$$

$$\sigma_d = \sigma_c \left[\frac{3}{4}\left(\frac{h}{t}\right)^2 + 1\right], \quad \text{(Eq. 2)}$$

where $h$ is the height of the buckle, $w$ is its width, $t$ is the thickness of the film, $v$ is its Poisson's ratio and $E$ is its elastic modulus.

Finally, the interfacial fracture energy ($\gamma$) for spontaneous buckles is expressed as

$$\gamma = \left[\frac{(1-v^2)t}{2E}\right](\sigma_d - \sigma_c)(\sigma_d + 3\sigma_c). \quad \text{(Eq. 3)}$$

CVD of $h$-BN on graphene film results in compressive strain (Fig. 4 in the main manuscript), and we observe the formation of spontaneous buckles when $t > 50$ nm (Fig. S12a). We measured the geometry of the spontaneous buckles by atomic force microscopy (AFM) and used the measured values to deduce $\gamma$ by using Eq. 3. A height profile (Fig. S12b) of a buckle measured by AFM yielded $h = 440$ nm, $w = 3.6$ μm and $t = 60$ nm; estimated $\gamma = 7$ meV per carbon atom. ($E$ and $v$ of $h$-BN film are 865 GPa and 0.211, respectively according to a previous report.[10]) This value is somewhat lower than the estimated interfacial adhesion energy of 23 meV/C atom that was calculated from the accumulated critical strain energy in the graphene film prior to the buckling (main manuscript). We assume that the strain energy is fully released by isolation of the film, and becomes equivalent to $\gamma$. In reality, some energy will remain in the final film, so 23 meV/C atom can be considered as an upper limit of $\gamma_{Gr-Ge}$.

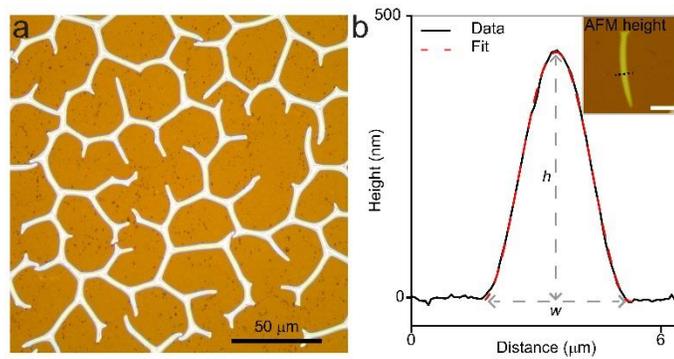

**Figure S12** (a) Optical image of *h*-BN/graphene film on a Ge(110) substrate after spontaneous buckling of *h*-BN/graphene film that was 60 nm thick. (b) Height profile measure across a buckle shown in the inset AFM image. Scale bar: 10 μm.

**Interlayer optical absorption peak area in artificially fabricated tBLG**

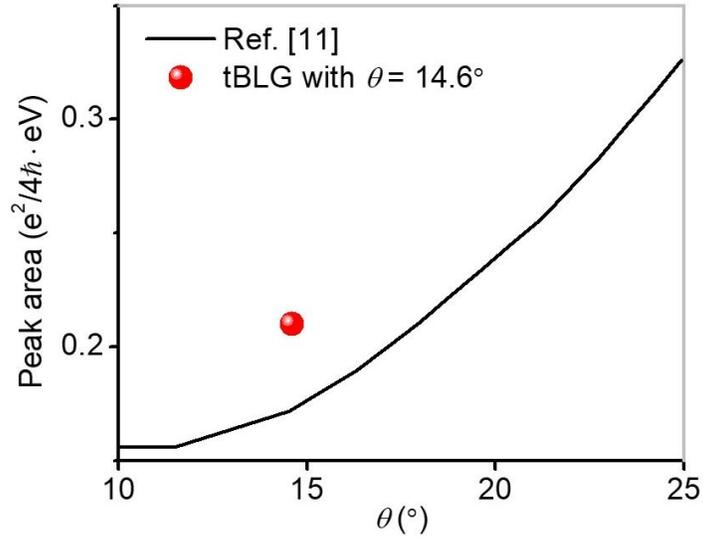

**Figure S13** Interlayer optical absorption peak area as a function of interlayer rotational angle ($\theta$). Interlayer optical absorption peak is determined by a Gaussian fit of the $\sigma_{tBLG} - 2 \times \sigma_{SLG}$ spectra. (main manuscript, Fig. 5c) The black curve is taken from a reference[11], and the red circle indicates the value measured from our tBLG sample. The $\theta$ is estimated to be 14.6° on basis of the measured interlayer optical absorption peak energy. The integrated absorption peak intensity is comparable to the reference value, suggesting a formation of clean interface with efficient interlayer interactions.